# Observations of cyclone-induced storm surge in coastal Bangladesh

Soyee Chiu[1*], Christopher Small[1]

[1]*Lamont-Doherty Earth Observatory, Columbia University, Palisades, NY 10964, USA*
[*]*sc3245@columbia.edu*

**Abstract**
Water level measurements from 15 tide gauges in the coastal zone of Bangladesh are analyzed in conjunction with cyclone tracks and wind speed data for 54 cyclones between 1977 and 2010. Storm surge magnitude is inferred from residual water levels computed by subtracting modeled astronomical tides from observed water levels at each station. Observed residual water levels are generally smaller than reported storm surge levels for cyclones where both are available, and many cyclones produce no obvious residual at all. Both maximum and minimum residual water levels are higher for west-landing cyclones producing onshore winds and generally diminish for cyclones making landfall on the Bangladesh coast or eastward producing offshore winds. Water levels observed during cyclones are generally more strongly influenced by tidal phase and amplitude than by storm surge alone. In only 7 of the 15 stations does the highest plausible observed water level coincide with a cyclone. While cyclone-coincident residual water level maxima occur at a wide range of tidal phases, very few coincide with high spring tides. Comparisons of cyclone-related casualties with maximum wind speed, hour of landfall, population density and residual water level (inferred storm surge) show no significant correlations for any single characteristic. Cyclones with high casualties are often extreme in one or more of these characteristics but there appears to be no single extreme characteristic shared by all high casualty cyclones.

Additional Index Words: *tropical cyclones, Ganges-Brahmaputra delta, tide gauges, casualties, tide network, storm surge, coastal impacts.*

**Introduction**
Tropical cyclones in the Bay of Bengal (BoB) have a large impact on the coast of the Ganges-Brahmaputra delta (GBD). High winds combined with high water levels cause extensive property damage and loss of life worldwide, but particularly in this densely populated coastal region (Nicholls, 2006; Nicholls, Mimura and Topping, 1995). The magnitude and impact of storm surges are of particular interest. Storm surge is defined by the National Oceanic and Atmospheric Administration's (NOAA) National Hurricane Center as the "abnormal rise of water generated by a storm, over and above the predicted astronomical tides" (www.nhc.noaa.gov/surge). Despite great interest and a considerable body of literature, most studies of storm surge on the Bangladesh coast focus on impacts of individual cyclones or on the modeling of cyclones and storm surges. We are not aware of any comparative studies of cyclones and storm surges using water level observations on the Bangladesh coast of the GBD. The objective of this study is to quantify the systematics of cyclone-driven storm surge on the Bangladesh coast of the GBD using cyclone characteristics (wind speed, size, proximity) and observations of water level as measured by a network of tide gauges.

JTWC Cyclone Tracks, 1977-2013

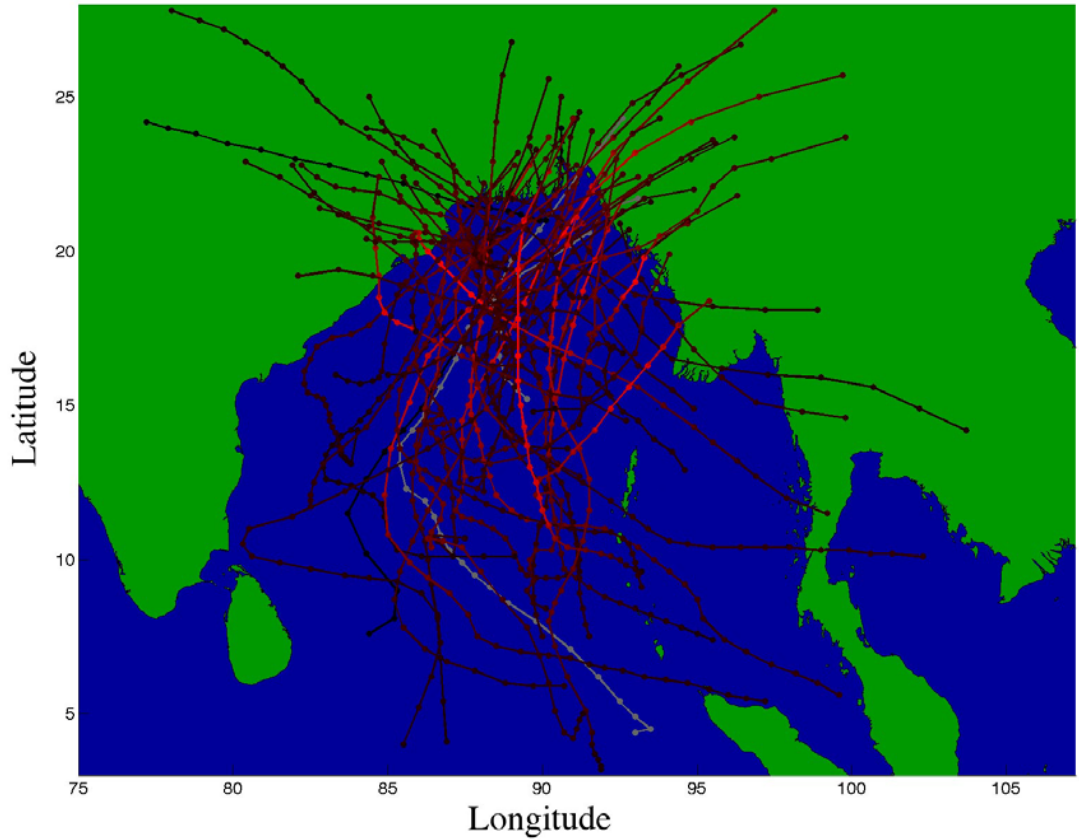

Figure 1. Tracks of cyclones in the northern Bay of Bengal between 1977 and 2013. Tracks shown in grey occurred in and after 2011 for which we do not have water level data. Filled circles correspond to six-hour intervals for which wind speed estimates are available. Landfall density is nearly uniform with longitude in the northern bay with high wind speed landfalls occurring over the range of longitudes. Wind speed alone does not determine depth of the cyclones' penetration on land and the geographical extent of populations exposed to their effects.

In this analysis we use water level records from a network of 15 tide gauge stations on the lower GBD to quantify systematics of 54 cyclones occurring in the BoB between 1977 and 2010 (Figure 1). We infer storm surge magnitude by estimating residual water levels from observed water levels by modeling and subtracting the astronomical tide. Following the convention of Horsburgh and Wilson (2007), we use the term "surge" in reference to known or inferred meteorological contributions to local sea level. We refer to "residuals" as the difference between observed water levels and modeled tides with the recognition that residuals may still contain nonlinear tide-surge interactions (including contributions from frictional and shallow water effects), harmonic prediction errors and timing errors. We compare residuals for different cyclone landfall locations to infer the effect of cyclone proximity and wind direction on water levels and possible surge heights. For all major (high casualty) cyclones and a selection of other cyclones, we compare time series of recorded water levels, modeled astronomical tide and computed residuals for all stations with usable data (defined below and in the Methods section). Cyclone-coincident water levels are compared to normal tidal variations in water level to provide a context for the impact of storm surge that does not rely on modeling the tidal constituents. The residuals are used to evaluate effects of wind speed, cyclone size, tidal phase and landfall location on water levels and inferred storm surge on the Bangladesh coast. Finally, we compare casualty estimates with other cyclone characteristics for a subset of 33 cyclones for which standardized casualty estimates are available for the three countries (India, Bangladesh, and Myanmar) that lie within our region of study.

*Background*
Due to the vulnerable nature of the coastal GBD region, disaster risk reduction comprises a large portion of cyclone induced storm surge literature. These studies often examine a single major event such as the May 1985 cyclone (Siddique and Euosof, 1987), the April 1991 cyclone (Bern *et al.*, 1993; Chowdury *et al.*, 1993) , and cyclones Sidr (Paul, 2009) and Aila (Mallick, Rahaman and Vogt, 2011) and their affect on the region's infrastructure and population. Furthermore, many storm surge modeling studies such as those described below aim to improve cyclone warning systems that are a part of disaster risk reduction programs. The focus of these studies is often to provide recommendations for decision makers rather than present new data on storm surge associated with cyclones.

Storm surge modeling is the prevalent method for estimating storm surge and damage from historical cyclones as well as for forecasting future storm surge. Studies that focus on a few historical cyclones, such as the April 1991 and November 1970 cyclones (Flather, 1994) and seven select cyclones from 1970-2000 (Dube *et al.*, 2004) did not have water level data available to validate their results. These hydrodynamic and atmospheric models require wind stress forcing, which is either derived from cyclone track/forecast wind estimates (Flather, 1994) or idealized (Dube *et al.*, 2004). Idealized cyclones with assumed parameters are also used (Das, Sinha and Balasubramanyam, 1974; Flierl and Robinson, 1972) to estimate surges. Tide-surge interaction and data estimation/measurement quality have been cited as reasons for over- and underestimating "observed" surges. The importance of water level data is underscored by the overestimation of total water level in Flierl and Robinson (1972) by simply superimposing modeled storm surge on modeled astronomical tides from tide tables. First order interactions between tides and surges are discussed in detail by Horsburgh and Wilson (2007), including the more complex nature of these interactions in shallow estuaries. The primary effect of these interactions is to introduce a phase shift generally causing high water levels to coincide with rising tides and precede high tide by several hours.

Cyclone characteristics have been shown to affect the magnitudes of storm surges as simulated by storm surge models. For example, a surge asymmetry is induced by offshore winds that produce negative surges to the west of landfall and onshore winds that produce positive surges to the east of landfall (Peng, Xie and Pietrafesa, 2006). While the sensitivity of offshore winds was found to be greater than onshore winds, Peng et al recognized the variability of their model results with varying cyclone parameters and acknowledge that the incorporation of wind inflow angle observations from Hurricanes Charley and Isabel improved wind field reliability and resulted in sea surface elevations closer to observed elevations. Weisberg and Zheng (2006) case study on Tampa Bay storm surges finds that the direction as well as speed of a hurricane's approach influences simulated surge heights. Though their area of study differs from Peng et al, they agree fundamentally with Peng on the effect of offshore and onshore winds producing negative and positive surges. Both also agree that other factors such as the hurricane's forward speed and intensity affect their modeled surge heights. It is important to note the small tidal range relative to surge height in Weisberg and Zheng's study area in contrast to the larger tidal range in the Bay of Bengal, for this influences the tide-surge interaction introduced above. Finally, Flierl and Robinson (1972) speak specifically about the Bay of Bengal, finding that a right angle curvature of the coastline, similar to the northeastern corner of the Bay of Bengal, produces almost double the surge than a straight coastline.

Surge modeling is also used for flood forecasting (Madsen and Jakobsen, 2004) and to predict extreme water levels (Jain *et al.*, 2010). In Madsen and Jakobsen (2004), observed air pressure and wind speeds are incorporated into a statistical cyclone track model which provide forecasts of surface air pressure and wind fields as input for a hydrodynamic model. The maximum simulated water levels differ as much as 4 m from observed values. Jain uses synthesized (combined observed and modeled) cyclone tracks from the east coast of India to predict maximum extreme water levels at a 50-year return period. Twenty-four tracks are used, with 10 located within our study region (Orissa and West Bengal state). While 15 days of tidal data were available to validate their tidal model, a comparison of observed water levels with surges from historical cyclone tracks was not included in their analysis.

Though models differ in forcing equations, assumptions, and implementation, they generally show sensitivity to cyclone track and wind speed data (*e.g.,* (Flather, 1994; Madsen and Jakobsen, 2004)). Thus forecast accuracy, both in time of landfall and magnitude of storm surge, are also limited by data available ahead of a cyclone's landfall. Using a network of tide gauges to compare water levels with historical cyclone data can provide insight to the consistency of tide gauges for analysis of short term events and can be useful for assessing the efficacy of these model forecasts when estimates of historical storm surges are compared.

There are few studies that focus on observations of storm surge in the BoB, especially in analyzing water level data in conjunction with cyclone characteristics. An early report of storm surges from four cyclones in 1960-1961 uses largely estimates and contain only two recorded water levels (Dunn, 1962). Analysis of storm surges has been done using water level measurements for three cyclones along the east coast of India (Sundar,

Shankar and Shetye, 1999). In this study, meteorological data was used in lieu of cyclone track data to identify significant cyclone events. The modeled residuals were found to be less than 100 cm with one of the three cyclones producing a run-down in water level rather than a positive surge. Lee (2013) identifies storm surges by applying Ensemble Empirical Mode Decomposition (EEMD) and Extreme Value Analysis (EVA) of water levels from a single tide gauge in Bangladesh. (Menendez and Woodworth, 2010) conduct a quasi-global statistical study using EVA on changes in extreme water levels in general as a function of seasonality, interannual variability (*e.g.,* El Niño) and long-term trends. Both of these studies focus on long term (multi-decadal) changes in extreme sea level rather than the relationship between individual cyclones and surges. A climatological perspective on tropical cyclone tracks in the BoB is offered in Islam and Peterson (2009), but without comparison to water levels. These studies focus on either a small (2-4) number of cyclones in the BoB or a large number of cyclones and their relationship to long-term sea level trends. To our knowledge, there are no studies of cyclones on the GBD coastline that compare characteristics of several cyclones and their associated surges using water level data from tide gauges or other direct measurements.

*Data*
The water level data consist of 34-year time series (1977-2011) obtained from 15 tide gauges throughout the Bangladesh coastal zone of the GBD (Figure 2). These float gauges are maintained by Bangladesh Inland Water Transportation Authority (BIWTA). There are three important considerations when using these data. First, because many of the gauges are located in remote areas of the coastal zone, we do not assume regular benchmarking and datum checks have been done. Second, we assume the datum used is a Chart Datum modified from the Indian Spring Low Water (ISLW) Chart Datum (Mondal, 2001) but we are unable to confirm this from the source of the data. The ISLW was intended for semi-diurnal dominated tides with small shallow water constituents. The additional term in the Chart Datum presumably takes into account these shallow water constituents, which can play a large role in tidal propagation in the Bangladesh coastal zone. Third, gaps in the time series exist where there are gauge malfunctions and reduced sampling rates (2 hourly or fewer, as well as gauges not recording in the evening hours). There are also notable vertical shifts in the time series at some stations. These data are not included in our analysis. Despite these limitations, we are able to quantify consistencies in cyclone-coincident water levels by using short (~1 month) intervals of continuous, non-shifted data in our analysis. No evidence is seen of the cyclones themselves impacting the gauges although some gauges have data gaps during some cyclones. We do observe cases where the gauge ceased operation during a cyclone and resumed afterwards, suggesting operators left the station. Because we cannot confirm that different gauges have been cross calibrated and are actually using the same datum, we do not directly compare water level measurements among gauges. We show measured water levels and residual heights for all gauges (with usable data) to provide an indication of consistency of response within and among cyclones at different locations. Hence, we assume that the measurements made by each gauge for each cyclone provide an accurate indication of the absolute change in water level before, during and after the cyclone but we assume nothing about the absolute differences of water level among different gauges.

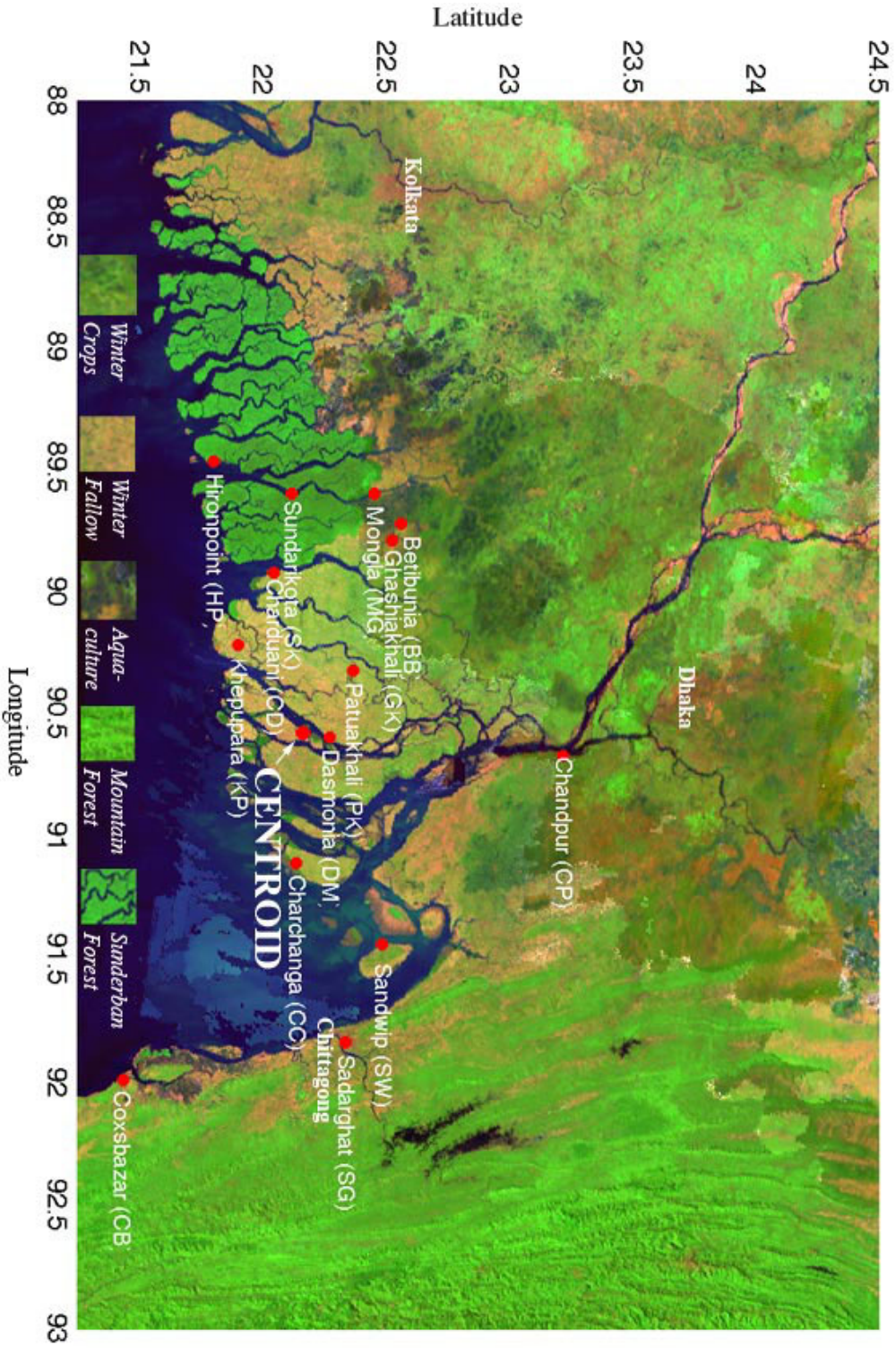

Figure 2. Tide gauge location map. Teknaf (located south of map extent) omitted for clarity. Centroid represents geographical mean of tide gauge locations. MODIS false color composite from 9 January, 2014 shows land cover types.

Cyclone track data, including six-hourly cyclone center locations and maximum (one minute sustained 10m) wind speeds, are provided by the Joint Typhoon Warning Center (JTWC) best track archive. The cyclones we analyze occurred in and after 1977, and their best tracks have been cross-validated with JTWC's Annual Tropical Cyclone Reports as well as a large observational network (Chu and Sampson, 2002). Wind speeds are estimated from the Dvorak method (Chu and Sampson, 2002). While the Indian Meteorological Department records cyclone best tracks with three minute maximum sustained wind speeds estimated with surface and upper air data, their time period is limited to post-1990s. A comparison of IMD and JTWC maximum wind speeds for overlapping JTWC/IMD cyclones yields reasonable agreement, with a best fit line of JTWC = 0.86 * IMD + 3.29 with r = 0.94. Systematic differences do occur between agencies recording cyclone tracks (Knapp and Kruk, 2010), but sufficient methodology documentation is needed to better quantify and explain these differences. Because the water level data are from a coastal region on a broad continental shelf, the effect of surface air pressure deficit is small compared to the effect of wind stress; thus we do not include surface pressure as a primary characteristic in our analysis.

According to the JTWC best track archive, there are a total of 56 cyclones that made landfall in northern Bay of Bengal between 1977 and 2013 (Figure 1). Water level data are available for 54 of these cyclones. Duration of the track data vary from 18 hours to 276 hours (11.5 days). Cyclone intensity, estimated by maximum wind speeds, varies from 50 km/h to over 250 km/h.

Population density estimates for the coastal zones surrounding the BoB are derived from gridded census enumerations compiled in the beta release of the Gridded Population of the World version 4 (GPW4) provided by the NASA Socioeconomic Data and Applications Center (SEDAC) (http://sedac.ciesin.columbia.edu/data/collection/gpw-v3). Maps of stable lighted development and inter-annual night light change are derived from the DMSP-OLS and VIIRS day-night band products provided by the NOAA National Geophysical Data Center Earth Observation group (NGDC EOG) (http://ngdc.noaa.gov/eog). Inter-annual night light changes are derived from intercalibrated OLS imagery and fused with higher spatial resolution VIIRS imagery following the method described by Small (2014).

Casualty estimates from cyclone events are a widely used metric to assess the severity of a cyclone and its impact on humans. Casualty estimates in previous studies came from a variety of sources, including local colleges and programs (Bern *et al.*, 1993), local statistical bureaus and disaster relief organizations (Siddique and Euosof, 1987), and local meteorological departments (Mallick, Rahaman and Vogt, 2011). These data are unstandardized across sources, and thus may lead to discrepancies due to different collection and recording methods. We use the Center for Research on the Epidemiology of Disasters' (CRED) Emergency Events Database (EM-DAT), whose data sources include the United Nations, national governments, and inter-governmental organizations. Most importantly, EM-DAT has an established recording method, which provides consistency in translating raw numbers of differing units (*e.g.,* one family in a developed country = 3 persons, one family in a developing country = 5 persons) to one database.

Casualty numbers are organized by cyclone and affected countries, and casualty totals for each cyclone are determined by the date(s) and location(s) of occurrence. Casualty data are available for 33 of the 54 cyclones we study. Damage in USD is also a popular metric for cyclone intensity. However, we choose not to include it in our analysis because although "several institutions have developed methodologies to quantify these losses in their specific domain […] there is no standard procedure to determine a global figure for economic impact." (Guha-Sapir, Below and Hoyois, 2014)

## Methods

From the BIWTA tide gauge records, we first select intervals spanning the duration of the 54 JTWC cyclone tracks. For each cyclone, tide stations with sampling rates < 2 hourly are discarded, as are stations with significant (> 2 days) gaps. Next, a classic harmonic tide model, *t_tide* (Pawlowicz, Beardsley and Lentz, 2002), was used to model the astronomical tide and compute residuals at each station during each of the 54 cyclones. The Raleigh resolution limit **$(N\Delta t)^{-1}$** (where N = number of data points, **$\Delta t$** = sampling timestep) is used to choose from all 45 astronomical and 24 major shallow water constituents. A range of 17 to 35 constituents are used. We then compute the residual by subtracting the modeled tide from the water level recorded by the tide gauge. Then we examine these residuals and adjust the water level intervals used to model to ~36 days surrounding the peak visible surge for all visible cyclones. This ensures that we capture the cyclones' full effect on water levels and also accounts for the water body's response time to the atmospheric forcing from the cyclone event. For those cyclones that did not display an obvious surge in the water level data, we used the JTWC best track maximum wind as a proxy "cyclone peak". We thought this reasonable because we observe peak storm surges coinciding with or occurring just prior to maximum wind speeds.

Highest plausible water levels for each tide station were identified after flagging isolated spikes of single high water level measurements showing implausible discontinuity with preceding and following hourly measurements that occurred during the period of available tide data. None of the flagged implausible measurements coincide with cyclones. Maximum cyclone water level residuals for each station were taken from subintervals spanning the duration of the cyclone determined by JTWC best track and storm surge analysis described above.

Spring high tides were identified using the Matlab local maxima function *findpeaks*, which defines local maxima as a point larger than its two adjacent points or any point equal to infinity. Spring tides were constrained to occur at most once every 300 hours (12.5 days) with water level greater than the station's mean, with the exception of the Chandpur gauge, which is heavily river discharge dominated and contains spring tides greater than and less than its mean.

Cyclone proximity was calculated as the great circle distance between the centroid of the tide station network and the point of landfall for each cyclone. Assigning negative distances for west-landing cyclones and positive distances for east-landing cyclones gives an indication of whether the region would be expected to experience primarily onshore or offshore winds from the cyclone and of the relationship between wind direction and storm surge magnitude.

## Results

Comparisons of measured water levels, modeled tidal constituents and residual heights (water level – tidal height) are shown for 8 high impact cyclones and for an additional 12 cyclones in Figure 3. For each cyclone, time series of all stations with usable data are shown for a full fortnightly tidal cycle to provide context for the normal water level fluctuations preceding and following passage of the cyclone. Wind speed estimates from the JTWC database are superimposed to provide a temporal context for the evolution and movement of the cyclone. It is immediately apparent that many of the cyclones do not coincide with any obvious change in water level distinct from tidal fluctuations. Despite the unknown intercalibration among tide gauges, the recorded water levels show generally consistent responses for the cyclones that do coincide with perturbations to the dominant tidal constituents. This provides some reassurance that the residuals that are observed (or not) are related to the cyclones and not other causes.

A comparison of residual height and location of cyclone landfall is shown in Figure 4. There is a clear and progressive change in residual height from west-landing to east-landing cyclones. Minimum residual height decreases monotonically with distance from the tide gauge network centroid from west to east with all cyclones. Both mean and minimum residual height decrease from west to east while maximum residual height reaches a maximum of ~2.6 m for cyclones Sidr and Aila making landfall ~160 km west of the centroid and diminishes monotonically in both directions. The negative residuals seen in the major April 1991 cyclone are recorded by tide gauges located west of landfall. This pattern is consistent with the expectation that west-landing cyclones should produce onshore winds at the longitude of the network and east-landing cyclones should produce offshore winds.

A comparison of maximum cyclone-coincident water levels from each tide station is shown in Figure 5. Red open circles indicate maximum plausible water levels, and blue dots represent maximum water levels achieved by each cyclone. Although not all tide gauges recorded usable data during every cyclone, we see a range of maximum cyclone-coincident water levels that only once falls outside of ~2.5 standard deviations of high spring tides (shown by whiskers). In only 7 of the 15 stations does the maximum plausible observed water level coincide with a cyclone.

Several cyclones on the BoB have made landfall on the densely populated GBD but a larger number have made landfall to the east or west of the most densely populated areas of the delta. Population densities are higher on the Indian coastline west of the delta and lower on the Burmese coastline east of the delta while most of the high density areas on the GBD occur on the upper delta more than 200 km inland (Figure 6). However, the coastal population density of ~1000 persons/km$^2$ on the Bangladesh coast is considerably higher than the global median density of ~500 persons/km$^2$ estimated by (Small and Nicholls, 2003). A significant fraction of the Bangladesh coast is also comprised of the effectively unpopulated Sundarban forest preserve. While this intertidal mangrove forest may help to buffer the effect of storm surges on human populations, the low lying polders surrounding the forest are especially susceptible to surges that propagate up the river channels past the Sundarban forest (Auerbach *et al.*, 2015). Satellite observations of night light suggest much greater intensity of development on the upper delta (Figure 6) but this is partially a function of differences in electrification within the country. Changes in night light indicate that much of the development in recent decades has occurred on the periphery of the larger cities and not in the coastal zone – aside from the city of Chittagong.

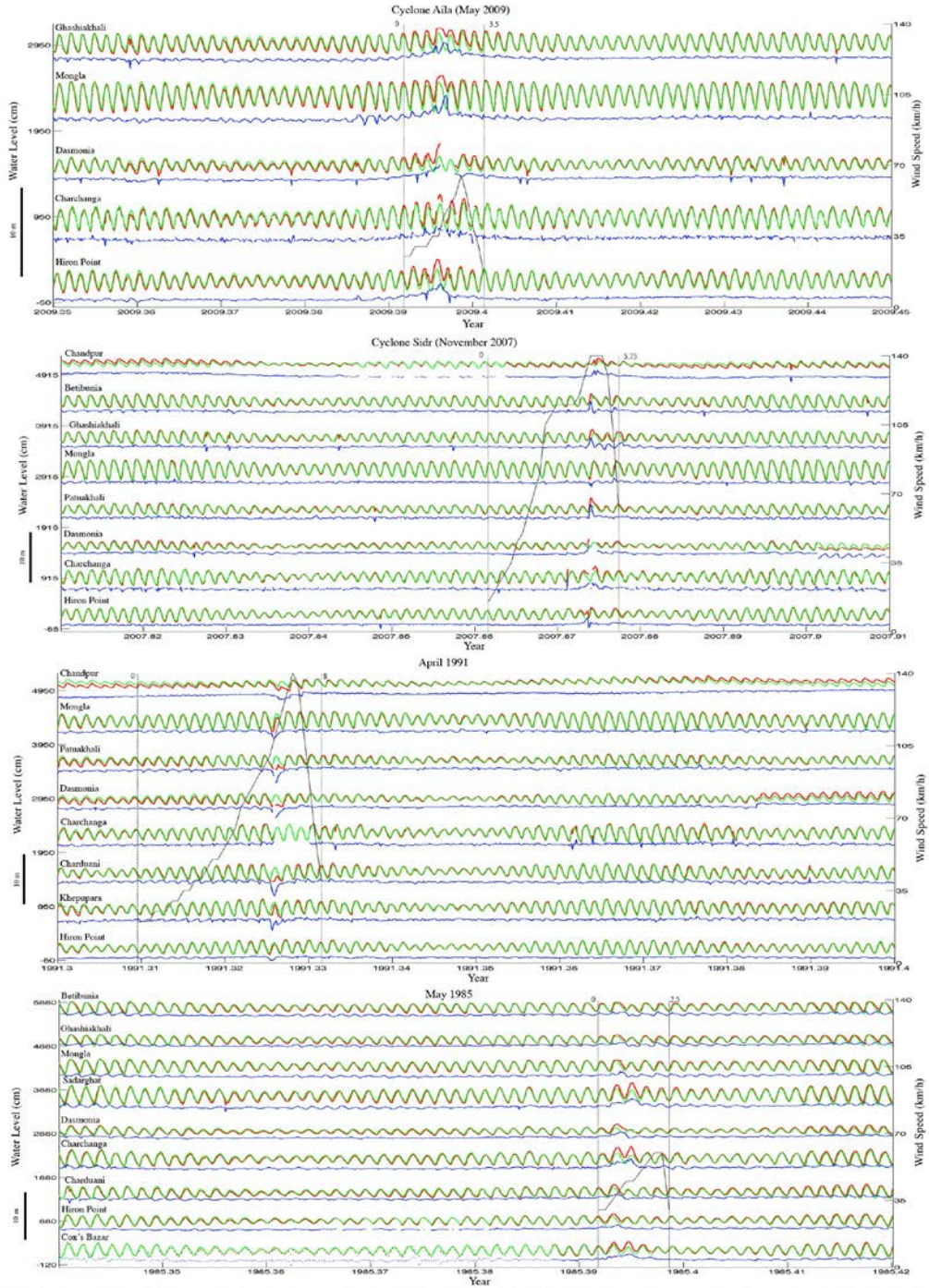

Figure 3. Water level records (red), modeled astronomical tides (green), water level residuals (blue), and wind speeds (black) for all 15 stations and 20 representative cyclones considered in this study. Not all stations recorded water levels for all cyclones. Stations are ordered by distance from the Bay of Bengal coast increasing from bottom to top. There appears to be no consistent relationship between water level and either wind speed or coastal proximity, though peak wind speeds in all but three cyclones occur after peak of residual. Although 13 of 20 cyclones appear to influence the water level, the large April 1991 cyclone coincides with a drop in water level.

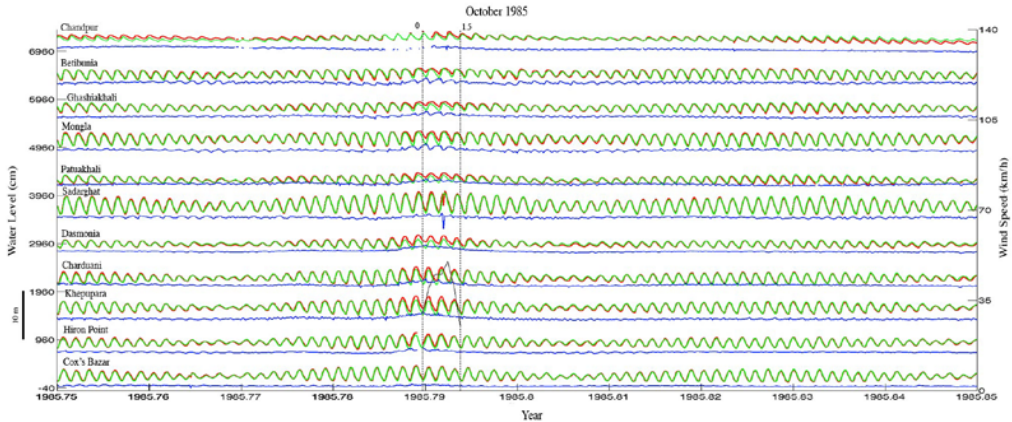
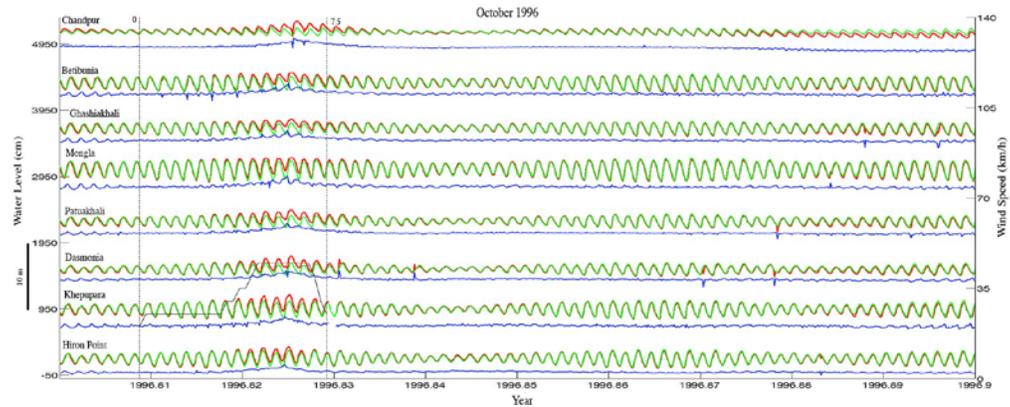
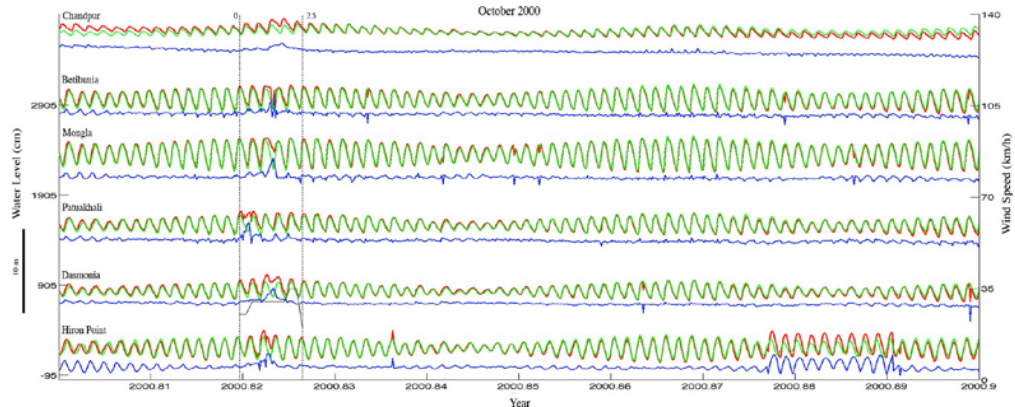
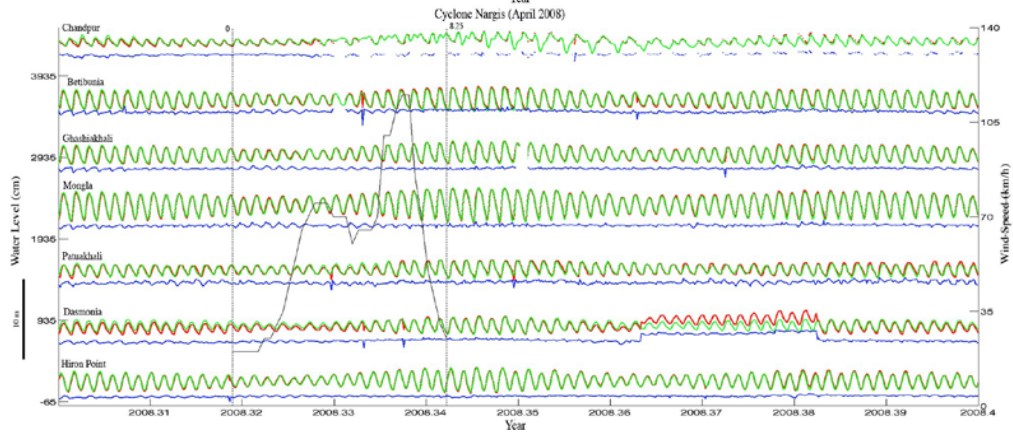

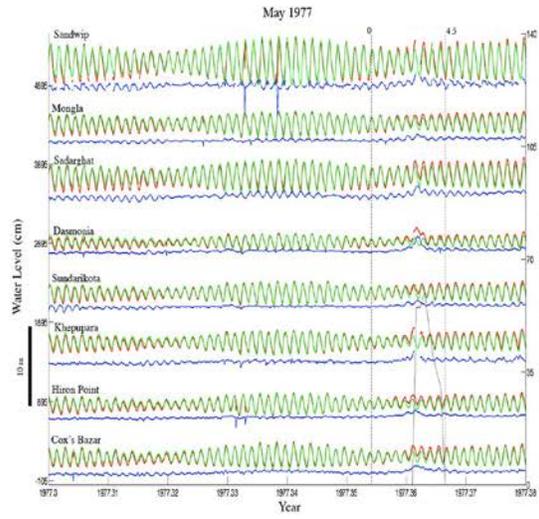
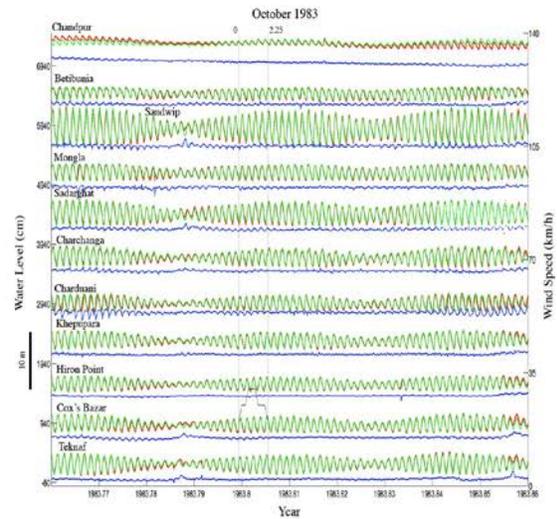
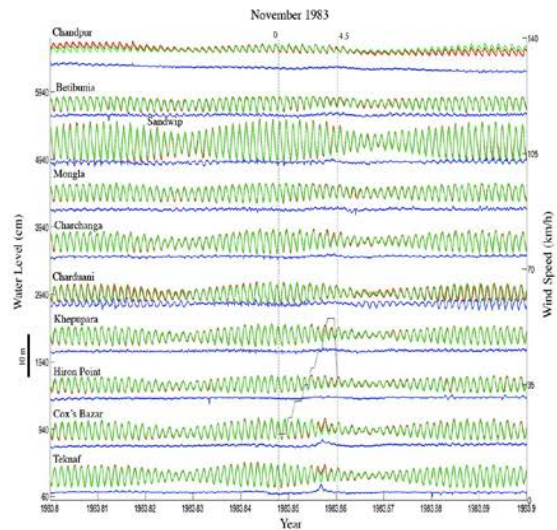
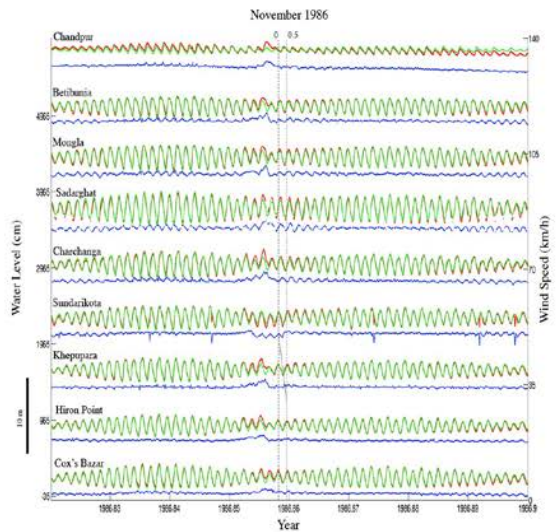
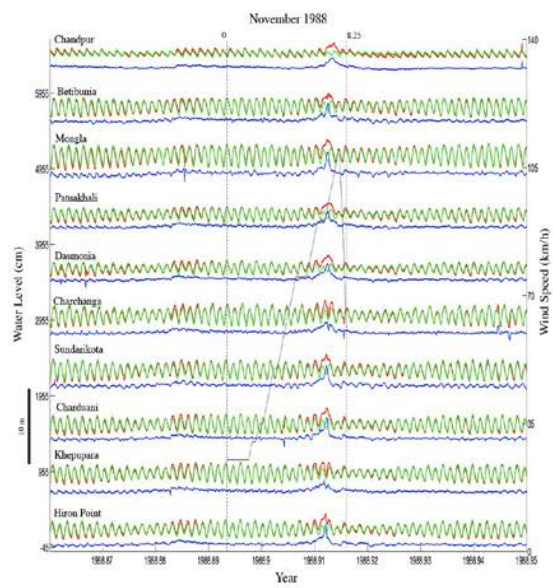
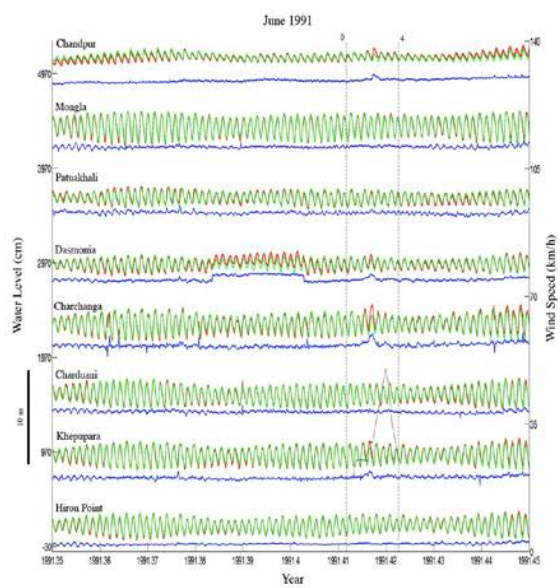

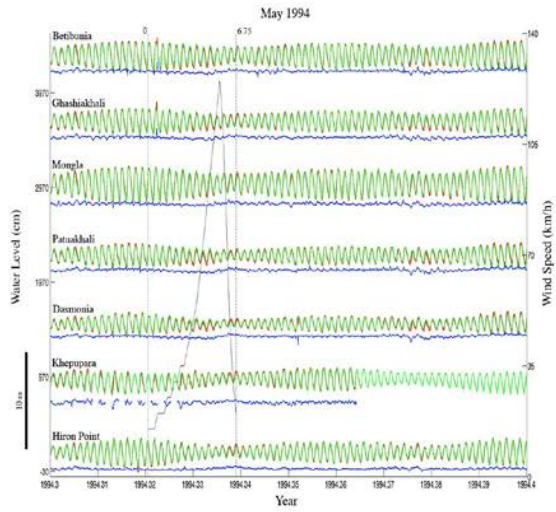
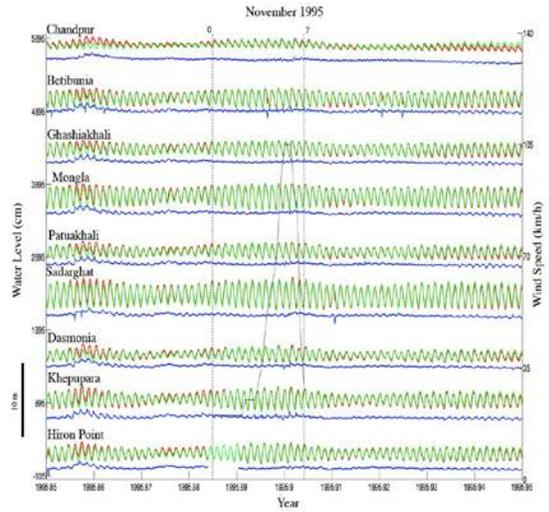
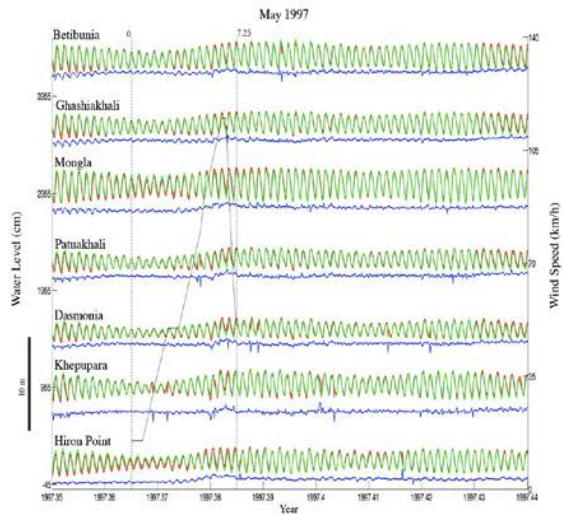
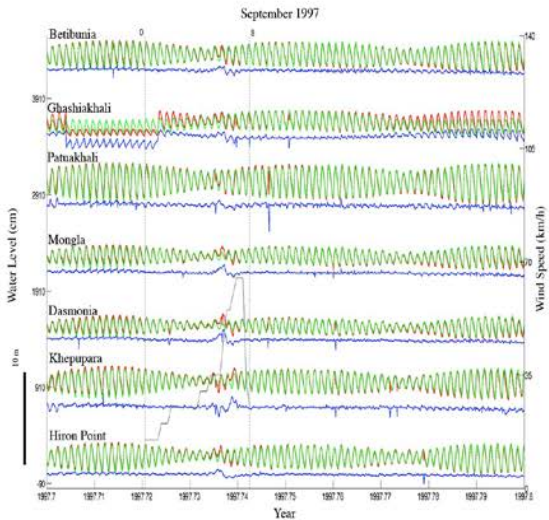
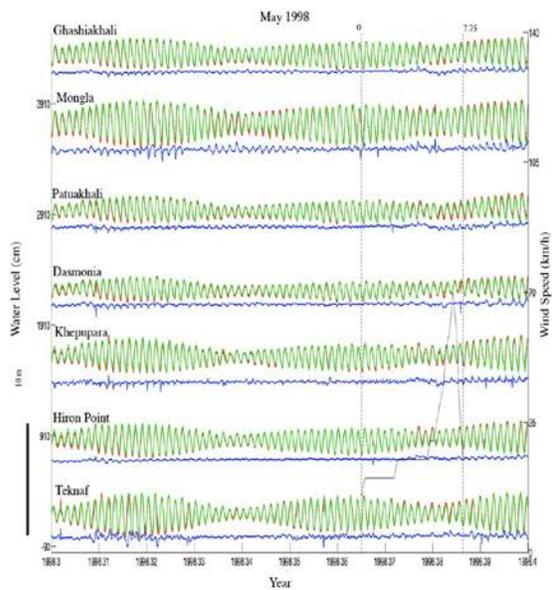
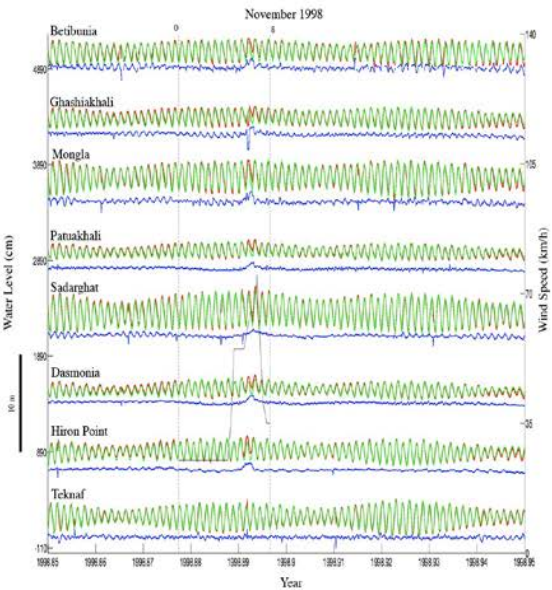

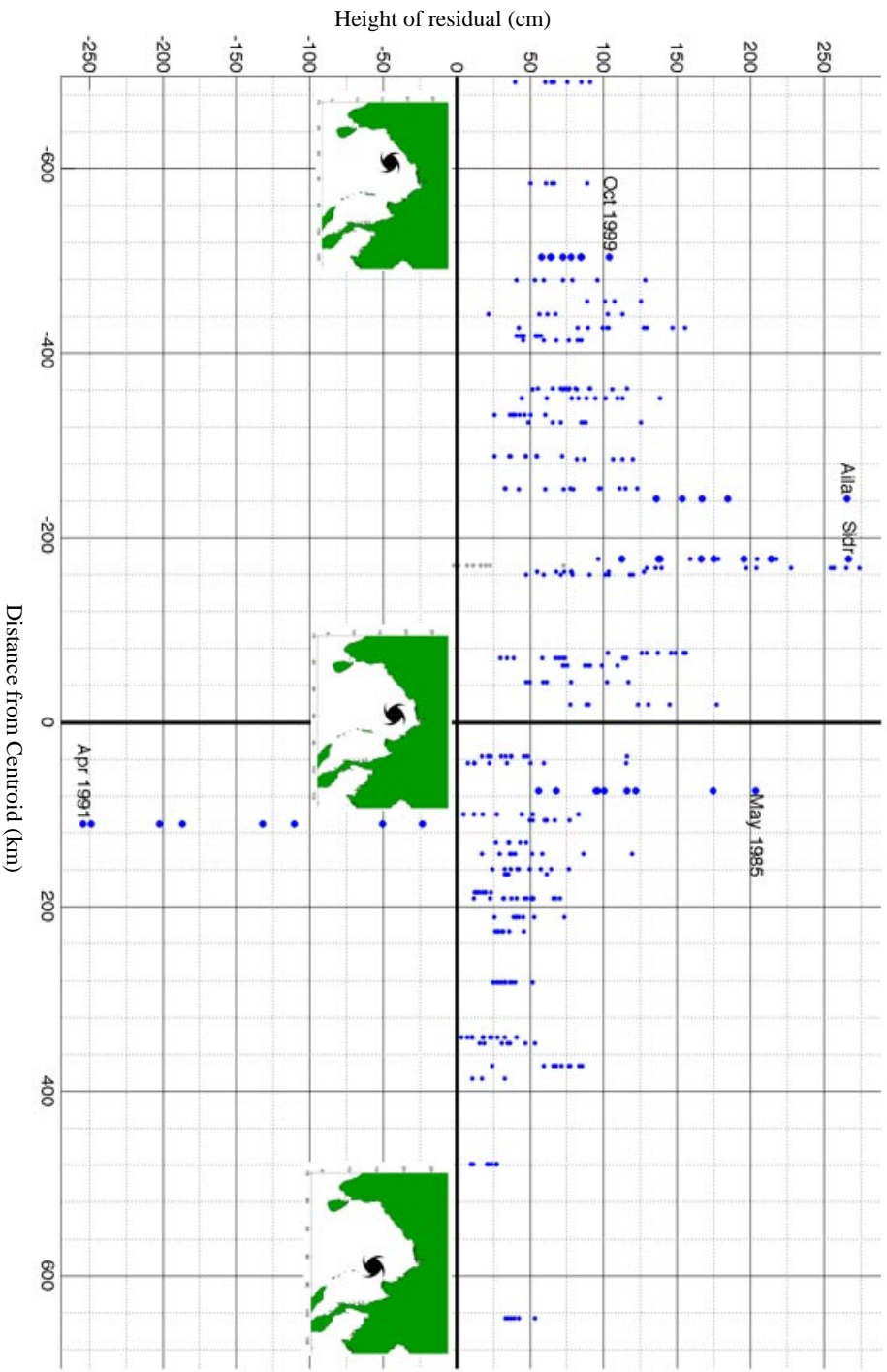

Figure 4. Residual and landfall. Distributions of residual magnitudes vary systematically with great circle distance from the tide gauge network. Cyclones making landfall west of the network generally result in higher maximum and minimum residuals than cyclones making landfall to the east. One exception is the September 1992 cyclone with smaller residuals (shown in grey), which made landfall during neap tide. This is consistent with the cyclonic winds being stronger onshore for west-landing cyclones and stronger offshore for east-landing cyclones. This is also consistent with offshore winds producing the run-down (negative residual) produced by the April 1991 cyclone.

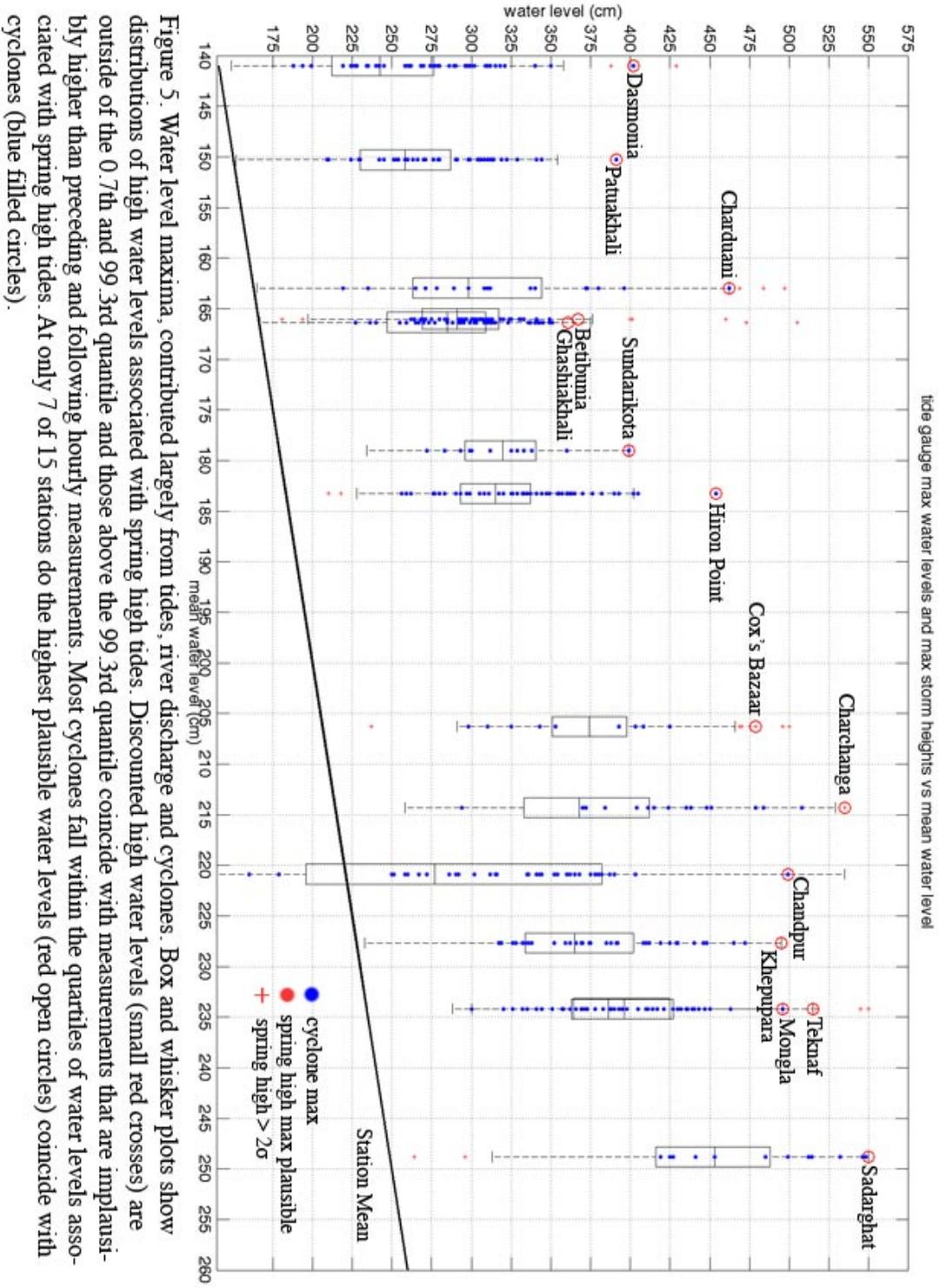

Figure 5. Water level maxima, contributed largely from tides, river discharge and cyclones. Box and whisker plots show distributions of high water levels associated with spring high tides. Discounted high water levels (small red crosses) are outside of the 0.7th and 99.3rd quantile and those above the 99.3rd quantile coincide with measurements that are implausibly higher than preceding and following hourly measurements. Most cyclones fall within the quartiles of water levels associated with spring high tides. At only 7 of 15 stations do the highest plausible water levels (red open circles) coincide with cyclones (blue filled circles).

We conclude the analysis with a comparison of casualty estimates and several characteristics of the cyclones for which there are comparable data. Figure 7 shows how casualties vary with wind speed, hour of landfall, residual height and population density. The casualty plots show no significant correlation for any of the characteristics, suggesting that no single characteristic is primarily responsible for cyclone casualties on the BoB.

## Discussion

Observed residuals computed from water level measurements on the BoB are generally smaller than reported surge heights. For example, our computed residuals are 0.06 to 5.03 m lower in 5 of the 6 cyclones with overlapping data studied by Dube *et al.* (2004). Though two of the six cyclones make landfall outside of the network of tide gauges that we use, the differences in reported surge and measured residuals are similar in magnitude to those cyclones making landfall within the network. While Dube *et al.* (2004) reported a 3.8 to 7 m surge for the April 1991 cyclone at stations east of the tide gauge network, we observe negative residual heights ranging from -0.23 to -2.52 m as well as clearly lower water levels in the tide gauge measurements at stations in the western part of the network. Three out of the four water level locations used by Dube *et al.* (2004) are in close proximity to the tide gauges used in our analysis. Similarly in Flather (1994), a positive surge of 5.2 m was reported with an error of ~2 m for the April 1991 cyclone. Despite the implausible outliers in the observed data (Figure 3), the observed residuals (and maximum water levels) at different stations are generally consistent in magnitude for each cyclone (Figure 3) and with landfall distance from the centroid of the network (Figure 4). We cannot comment with certainty on the accuracy of surge estimates published elsewhere because we have been unable to find any publication that describes how these surge estimates were obtained.

There is significant overlap between the range of observed residuals presented here and the range of modeled surge heights reported in earlier modeling studies. This suggests to us that the physical processes incorporated in the models are fundamentally correct. However, there is a conspicuous disparity between the larger reported surge heights (2 to 7 m) in several published studies and the largest residuals (< 3 m) observed in the tide gauge data. While we cannot independently verify the accuracy of the water levels measured by the tide gauges, we find the consistency across multiple gauges for almost all of the cyclone-coincident water levels reassuring. Moreover, the cyclone-coincident water levels reported here do not depend on the tide modeling component of the analysis and can be observed in the raw water level measurements (Figure 3). Nonetheless, both the measured water levels and computed residuals are inconsistent with the larger storm surge heights widely reported. The critical piece of information that seems to be missing is the process by which the surge heights in excess of 2 m were obtained. We assume that reported surge heights are based on *post hoc* field observations of inferred water levels rather than actual *in situ* water level measurements. If so, these observations are presumably given relative to some common elevation, like mean sea level. It seems unlikely that field observations of storm surge water levels are consistently referenced to the predicted astronomical tides at the time and location of impact – as implied by the definition of a storm surge. Therefore, we conjecture that reported storm surge heights may actually be storm tides, defined as "the water level rise due to the combination of storm surge and the astronomical tide" (www.nhc.noaa.gov/surge). However, several of the larger reported surge heights exceed even the full 2 to 3 m range (with the exception of Sandwip's 5 m range) of water levels observed in the tide gauge measurements (Figure 5).

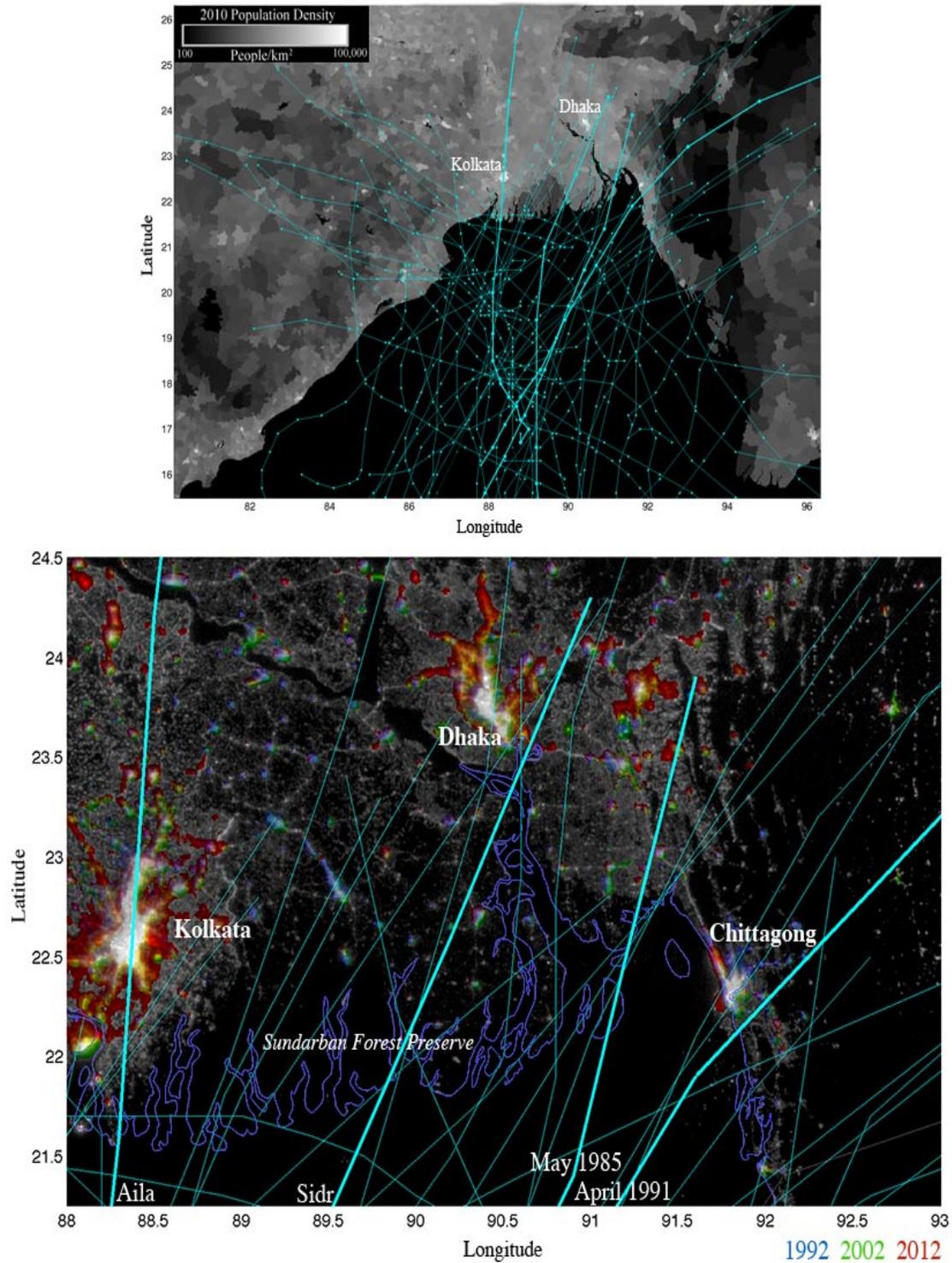

Figure 6. Storm tracks, population density and lighted development. While several cyclones have struck the densely populated Ganges-Brahmaputra delta (top center), the majority have made landfall in lower density areas to the east and west of the delta. Fused night light composite (bottom) shows decadal changes in luminance derived from DMSP-OLS annual composites fused with VIIRS mean composite from January 2013. Color implies change. Warmer colors show later brightening related to development and electrification. While the lower delta north of the Sundarban Forest Preserve shows less lighted development than the rest of the delta, rural population densities are comparable.

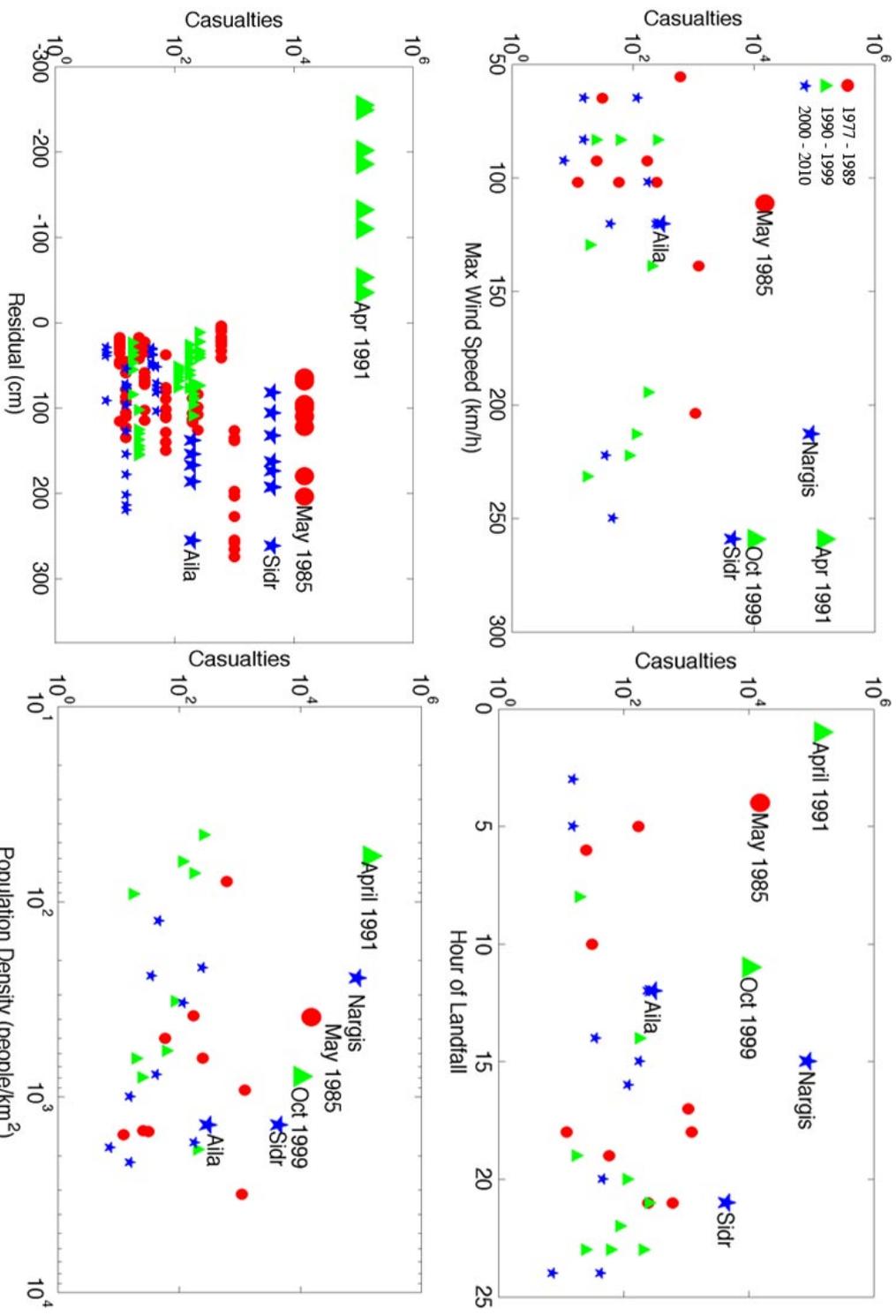

Figure 7. Cyclone characteristics and casualties. Casualties are included for Bangladesh, India, and Myanmar with the exception of the residuals panel, where only casualties from Bangladesh are represented. Cyclones with high casualties are often extreme in one or more characteristic but there appears to be no single extreme characteristic of all high casualty cyclones. For every characteristic, there are low casualty cyclones with similar values as high casualty cyclones. Each high casualty cyclone is different with respect to these parameters. For example, two of the highest casualty cyclones (May 1985 & April 1991) are both direct hits that made landfall in the middle of the night but have very different wind speed and residual.

There is a consistent difference between residuals from west-landing and east-landing cyclones (with respect to the centroid of the network) which coincides with the expectation that onshore winds drive higher water levels for west-landing cyclones and offshore winds drive lower water levels for east-landing cyclones. This is most apparent in the April 1991 cyclone (Figure 4), which made landfall east of all tide stations with usable data and resulted in significant negative residuals. In the May 1985 cyclone, stations to the west of landfall (offshore winds) exhibit a dip in residuals, while stations to the east (onshore winds) exhibit increased residuals. These examples further reinforce the consistency of the water level data. The progression shown in Figure 4 suggests that cyclones making landfall ~150 to 250 km west of the delta may be expected to produce larger storm surges than those making landfall near the center of the delta – although this is not always the case.

From the lack of correlation between cyclone characteristics and casualties (Figure 7), it appears that there is no single factor that is primarily responsible for high casualties among cyclones on the BoB. While some or all of these characteristics certainly contribute to casualties in some or all of the cyclones, there does not seem to be a dominant characteristic that is correlated with casualties in the majority of cyclones. Gridded census enumerations do show the highest regional population densities on the delta (Figure 6 top), but the highest population densities are largely inland and the cyclones with the largest numbers of casualties appear to show a diminishing trend with population density along their tracks (Figure 7 lower right). The populated coastal areas in Bangladesh have somewhat lower densities in comparison to inland areas, but by global standards they are still high at ~1000 people/km$^2$ (Small and Nicholls, 2003) Satellite observations of night light suggest significantly less lighted development on the lower delta (Figure 6 bottom), but the area surrounding the Sundarbans is known to have relatively high agrarian population densities. These high densities translate to increased risk of damage and loss of life from cyclone impacts. Because of the complex relationship of cyclone characteristics with casualties, it will be important to consider multiple characteristics in order to design accurate and effective guidelines for damage estimates/assessments as well as disaster risk reduction.

It has been proposed by Horsburgh and Wilson (2007) that residual water levels associated with storm surges on the North Sea generally precede high tides by 1 to 5 hours because water depth modulation of storm surge introduces a tidal phase shift that precludes the co-occurrence of storm surges with high tides. On the BoB, we observe that cyclone-coincident residual water level maxima occur at a wide range of tidal phases but very few coincide with high spring tides. This is consistent with the observations and hypothesis of Horsburgh and Wilson (2007). Although we observe no consistent clustering in phase as they observe, our sample is much smaller and perhaps not sufficient to establish robust distributions.

It has been proposed by Irish, Resio and Ratcliff (2008) that cyclone size plays an important role in surge generation. We attempt to test their model with cyclones on the BoB. Estimated radii of maximum wind are available in and after 2002, totaling 17 cyclones in the study area. Normalized residual water level is calculated from the mean residual water level divided by its pressure deficit (defined as the difference between far field surface pressure and the central pressure of the cyclone). Pressure and radii are taken at time of landfall using a far field surface pressure of 1020 mb as in Irish, Resio and Ratcliff (2008). The results show a weak correlation between cyclone size and residual water level (r=0.41). The 7 east-landing cyclones had a statistically significant ($p < 0.001$) correlation of 0.93 and the 10 west-landing cyclones had a statistically insignificant ($p \gg 0.05$) correlation of -0.05. While there is a general increase in minimum normalized surge with increasing radius, the highest normalized residuals (> 0.3 cm/mb) are associated with medium to small cyclones (28-55 km). We investigate this asymmetry in the relationship between cyclone size and residual height in greater detail in a separate study.

Although water level data for coastal Bangladesh show smaller residual heights (presumably indicative of smaller storm surges) than storm surges reported by previous studies, this does not discount the importance of storm surge on the area. On the embanked polders on the lower delta even a few tens of cm of above-normal water levels can overtop embankments and cause flooding, as occurred with cyclone Aila in 2009 (Auerbach *et al.*, 2015). The smaller residuals observed in water level observations do not in any way diminish the importance of storm surge in the lower delta, particularly for west-landing cyclones coinciding with spring high tides.

## Conclusions

Tropical cyclones are an important part of understanding coastal processes on the GBD. While many previous studies use models to estimate storm surge heights, we have computed residual heights from a network of observed water levels during and immediately surrounding 54 cyclones making landfall in the northern BoB between 1977 and 2010. Because we use a simple harmonic tidal analysis to detide the data, the resulting residuals may still contain nonlinear tide-surge interactions, harmonic prediction errors and timing errors. However, we find consistency in observed residuals across tide stations for the cyclone time intervals. The observed residuals are generally smaller than reported surge heights, and many cyclones show no obvious residual at all. This suggests that the phase of the 2+ m tide on the Bangladesh coast is a critical determinant for the impact of storm surges. Furthermore, there is consistency in residuals with respect to proximity to and direction from landfall location. West-landing cyclones, producing onshore winds on the Bangladesh coast result in generally higher residual water levels than east-landing cyclones producing offshore winds. While these results show smaller residuals than the storm surge heights reported in most previous studies, this does not in any way reduce the importance of storm surges in the Ganges-Brahmaputra delta. These results highlight the importance of landfall location and tidal phase to the development of storm surge on the coast of Bangladesh. The highly variable relationships between cyclone-related casualties and cyclone characteristics suggests that no single characteristic is primarily responsible for casualties on the BoB and casualties in any given cyclone may depend on a combination of cyclone characteristics – as well as other factors not included in this analysis (e.g. cyclone warning systems, cyclone shelter infrastructure).

## Acknowledgements

This research was supported by the Office of Naval Research (grant (N00014-11-1-0683)). The authors are grateful to Steve Goodbred, Dhiman Ranjan Mondal, Carol Wilson and five anonymous reviewers for helpful comments and suggestions.